\documentclass[a4paper]{jpconf}
\usepackage{graphicx}
\usepackage{amssymb, colortbl}

\begin{document}
\title{Newtorites in  bar detectors of gravitational wave}

\author{Francesco Ronga (ROG collaboration) \footnote{ROG collaboration: M. Bassan,E. Coccia, S. D'Antonio,V. Fafone, G. Giordano, A. Marini, Y. Minenkov,I. Modena,  G. V. Pallottino, G. Pizzella,A. Rocchi, F. Ronga, M. Visco.} }

\address{INFN Laboratori Nazionali di Frascati via Fermi, Frascati I 00044, IT} 

\ead{Francesco.Ronga@lnf.infn.it}

\begin{abstract}
The detection of  particles with only gravitational interactions (Newtorites)
in gravitational bar detectors
 was studied in 1984 by Bernard, De Rujula and Lautrup.
The negative results of dark matter searches suggest to look to exotic
possibilities like Newtorites.
The limits obtained with the Nautilus bar detector will be presented and
the possible
improvements will be discussed.  Since the gravitational coupling is very
weak, the possible limits are very far from what is needed for dark matter,
but for large masses are the best limits obtained on the Earth.
 An update of limits for {\it MACRO} particles will be given.
\end{abstract}

\section{Introduction}
Many experiments have searched  for supersymmetric  WIMP dark matter (DM), with null results. This  may
suggest to look for more exotic  possibilities.
 In this paper we will extend our previous  exotic particle  search  with gravitational wave ($gw$) cryogenic bar detectors \cite{Astone:2013bed} to the detection of particles having  DM with only  gravitational interactions. Newtorites were proposed   in 1984 by Bernard, De Rujula and Lautrup \cite{deru}. In this case the excitation of a bar detector is due directly to the newtonian force and the signals are very small because the newtonian force is extremely weak.
         We will focus on the Nautilus and Explorer detectors that our group operated for decades.
We will  describe the analysis procedure and the selection criteria to identify candidate events.
Then, starting from the energy distributions of the candidates events  we will present upper limits to Newtorites.
The same data can be used to  give  upper limits on a generic particle with strong interaction ({\it MACRO}) as suggested in  \cite{Jacobs:2015csa}. 

The sensitivity to Newtorites of  interferometric $gw$ wave detectors  will be also briefly discussed.

\begin{figure}
  \includegraphics[width=70mm]{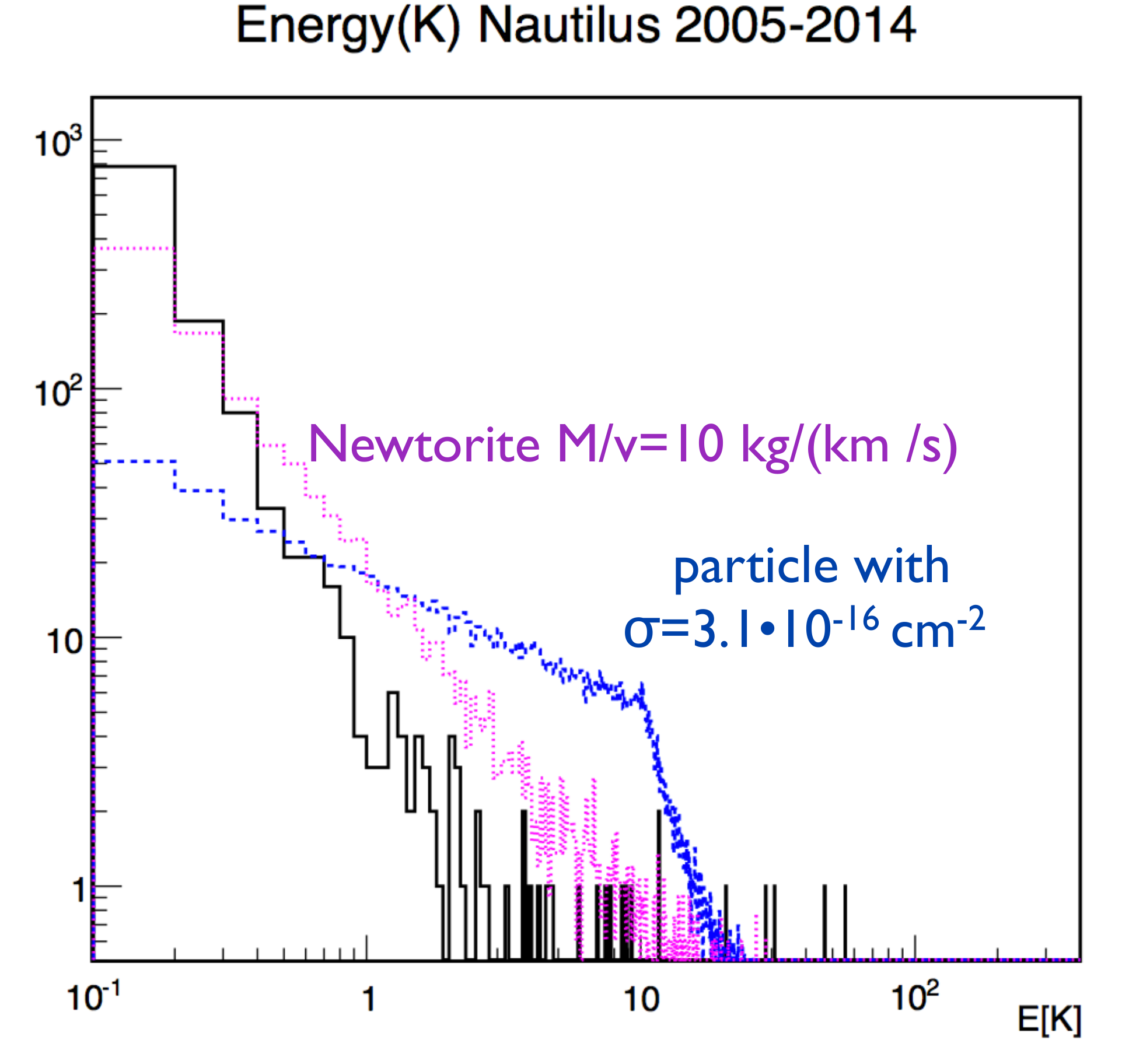}
\begin{minipage}[b]{9cm}\caption{\label{fig:Econfronto}Nautilus 2005-2014. Event energy distribution measured in Kelvin units (continuous line) compared with a Montecarlo for a particle having cross section $\sigma=3.1\cdot10^{-16}cm^2$  and $\beta=10^{-3}$ (dashed, blue online) and with the Newtorite
Montecarlo for $M/v$=10 kg~s/km (dots, magenta online). 
Montecarlo events are normalized in order to have the same number of events as the real data.}
\end{minipage}
\end{figure}

\section{The Nautilus and Explorer Data }
The gravitational wave  $gw$  detector Nautilus\cite{nautilus} is located in Frascati (Italy) National Laboratories of INFN.
Nautilus started  operations around 1998. The current  run started in 2003. The  detector Explorer,  similar to Nautilus,
was located in CERN (Geneva-CH). The Explorer run ended in June 2010. 

Both detectors use the same principles of operation.   Explorer and Nautilus consist of a large aluminum alloy cylinder (3 m long, 0.6 m diameter) suspended in vacuum by a cable around its central section to reduce the seismic and acoustic noise and cooled to about 2 K by means of a superfluid helium bath.
The ($gw$)  excites the odd longitudinal modes of the cylindrical bar.
  To record the vibrations of the bar first longitudinal mode, an auxiliary mechanical resonator tuned to the same frequency is bolted on one bar end face. This resonator is part of a capacitive electro-mechanical transducer that produces an electrical a.c. current that is proportional to the displacement between the secondary resonator and the bar end face. Such current is then amplified by means of a dcSQUID 
superconductive device and recorded on disk with an ADC sampled at 5kHz.
Both detectors are equipped with cosmic ray telescopes to veto excitations due to large showers.

 The data are processed off-line, applying adaptive  frequency domain filters optimized for short delta-like signals.
In order to select clean events  we have applied several cuts to the data.
The most important cuts are based on: the noise (average of the output in 10 minutes periods), the gain of the electronic chain, the SQUID  locking working point, 
the seismic monitors, the event shape. We have removed also periods with an high event rates and runs having live-time less than 10h.
The efficiency of those cuts is verified continuously using the extensive air showers detected by the cosmic ray detector.

 We have used only the Nautilus data to give Newtorite upper limits. We have used  the full data set, including Explorer,  to give  upper limits to {\it MACRO} particles.
The total live-time of the Nautilus  data set is 1846 days.
The energy distribution of the 931 Nautilus events surviving the cuts and having E$\ge$0.1~K is shown in Fig. \ref{fig:Econfronto}, together with the result of a Montecarlo simulation of a particle having  cross section $\sigma =3~10^{-16}$cm$^{-2}$  and $\beta=10^{-3}$ and the result of a Montecarlo simulation of a Newtorite having M/v=10 kg s / km (see later).
 
\section{Newtorite signal}
Due to the long range nature of the newtonian force, Newtorite signals could occur even if the particle
does not 
cross the bar. In the case of a  point like particle moving with a constant velocity $v$ along a
straight trajectory coming from infinity and going to infinity,  the vibrational amplitude of the
nth-vibrational mode is given by\cite{deru}:

\begin{figure}
 \includegraphics[width=70mm]{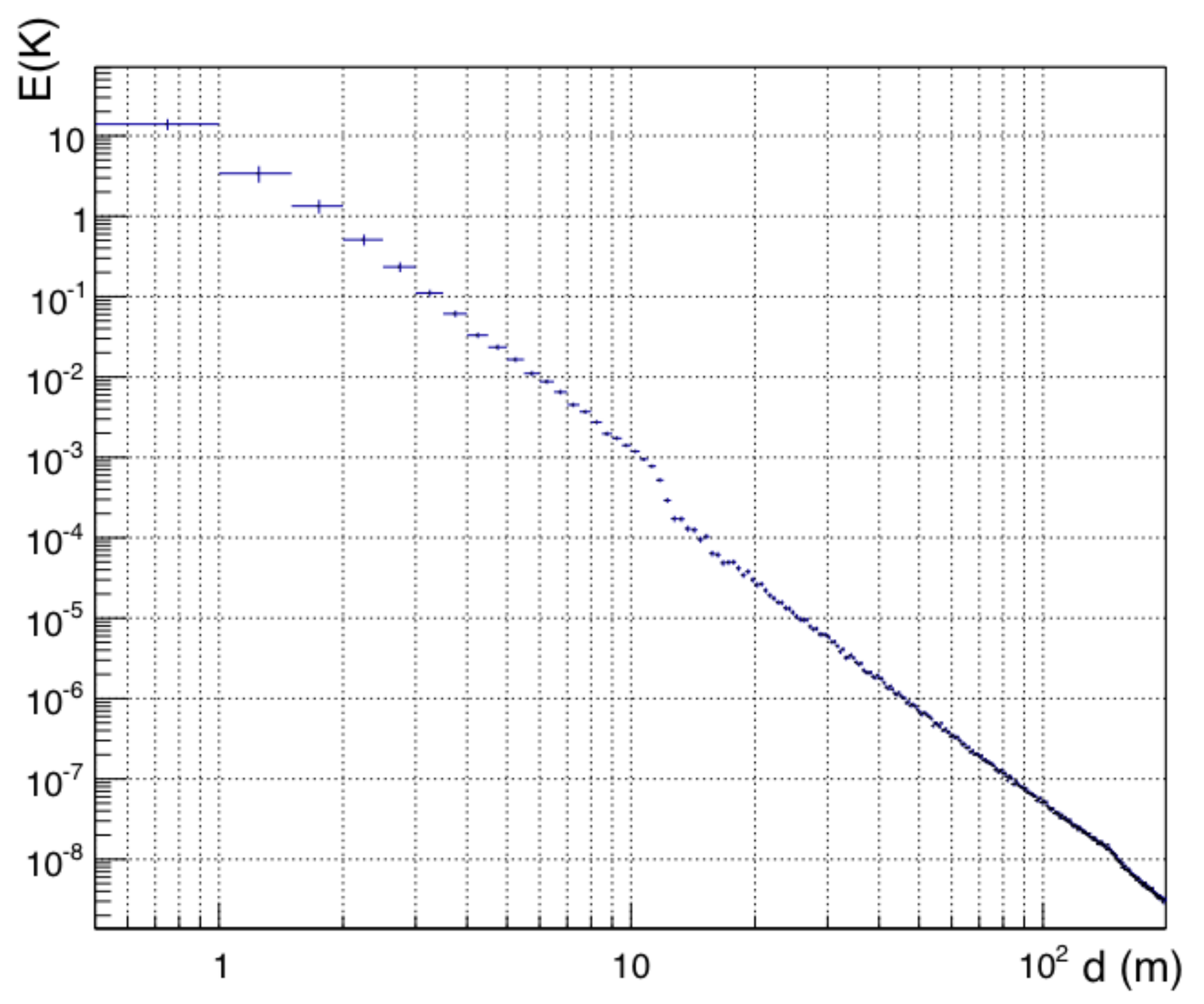}
 \begin{minipage}[b]{8cm}\caption{\label{figedist}Average from different directions of the signals due to a M/v=10 kg s / km Newtorite vs. distance from the bar center. The signal for different values of  $M/v$ (see eq.(\ref{eq:Tmiddle})), scales as $(M/v)^2$}
\end{minipage}
\end{figure}

\begin{equation}
A_{n}= -\frac {2GM}{Vv} \int_{V}   \frac { \mathbf {u_{n} \cdot }  
  \mathbf {(x_{T}- x_{T}^{0}) } }  { (x_{T}- x_{T}^{0}) ^2} d^3\mathbf{x}
\label{eq:Anewton}
\end{equation}
Here $G$
is the gravitational constant, M the mass of the bewtorite,
$\bf{x_{T}}$  are the transverse coordinates of a volume element of the detector relative to a fixed point $\bf{x_{T}^{0}}$, arbitrarily chosen along the particle track ; $\bf{u_n}$    is the spatial part of the nth
normal-mode oscillation normalized to the volume $V$ of the bar.
For a thin bar with radius $r$ and length $L$  ($r \ll L$)  
$\bf{u_n}$   can be approximately written, using cylindrical coordinates. as: 

\begin{eqnarray}
\label{eq:un}
& u^r_n= \sqrt 2 \sigma_{P} \pi (r/L) sin(n\pi z/L) \nonumber \\
& u^z_n= \sqrt2  cos(n\pi z/L)
\end{eqnarray}
Here $\sigma_{P}$ is the aluminium Poisson module.  The energy variation  in the bar is obtained by: 
\begin{equation}
\label{eq:Tnewton}
 \Delta E_{n}= \frac {1} {2 k_B } \rho A_{n}^2 V  \hskip1cm  [\rm{K}]
\end{equation}
Here {$k_B$} is the Boltzman constant.
In this paper we are only interested  in the first longitudinal mode n=1, and we assume
that the velocity $v$ of the particle is large enough  that most  of the signal is contained in a few ms.
This requirement is due to the $\delta$-like filter used to extract the antenna events.
Different filters could in principle detect longer signals. 
The signal is a fairly complicated function of the Newtorite's 
trajectory and has been computed in \cite{deru} in the particular case of orthogonal trajectory in the
middle of the bar,  and  for $r/L\to0$. In this case we can put  $u^r_1=0$ and we obtain:
\begin{equation}
\label{eq:Tmiddle}
 \Delta E \sim 30  \pi r^2 \frac{ \rho G^2 } {k_B  L} (\frac{M}{v})^2   \hskip1cm  [\rm{K}]
\end{equation}
Numerically we have for  Nautilus  $ \Delta E \sim  2.4 (M/v)^2$ K with $M$ expressed in kg and $v$ in
km/s.

In the general case the signal has been computed by numerical integration of eq.\ref{eq:Anewton}
inserted in a Montecarlo to simulate random directions. The result of one of those calculations as
a function of the distance of the trajectory from the bar center and for $M/v$=10  kg~s/km is shown in
Fig~\ref{figedist}. 
From this figure we can see that, at large distance d, the signal
energy scales as 1/d$^4$,  as expected from eq.\ref{eq:Anewton}. The signal falls below the energy threshold used in this analysis (see the next paragraphs) for d larger
than $\sim$ 3 m for $M/v$=10 kg~s/km. For $M/v$=100 kg~s/km this threshold occurs at $d \gtrsim$ 10 m.

 As a consequence, it is more important to select data with very low
noise than to increase the livetime. In Tab.~\ref{tb6}  we  show the limits obtained  using the Nautilus 2011 only. The limits with the full Nautilus data are typically a factor 2 higher.

\begin{table}
\centering
\begin{tabular}{|c	|c	|c	|c   |c }
\hline
 
M/v  &  acceptance 	&   events   &  flux upper  limit\\
 $kg~ km^{-1}~ s$      &   ($m^2 sr$) 	&  upper limit   &  $(cm^{-2} s^{-1} sr^{-1)}$ \\
 \hline
1          &33.4     &  37         &4.5 $\cdot 10^{-12 }$ \\
2          &85.3    &  23         &1.1 $\cdot 10^{-12}$ \\
5          &209     & 22	          &4.3 $\cdot 10^{-13}$ \\
10        &426     & 22	          & 2.1 $\cdot 10^{-13}$  \\
20        &888      &22	     	 & 1.0 $\cdot 10^{-13}$\\
40        &1652     & 22	    	 & 5.5 $\cdot 10^{-14}$\\
60         &2398    & 21	      	 & 3.5 $\cdot 10^{-14}$\\
\hline
\end{tabular}
\caption{Nautilus 2011. Newtorite acceptances and upper limits. Livetime=278.8 days, 160 events $\ge$  0.1K}
\label{tb6}
\end{table}
 
\section{Newtorite and MACRO limits and possible improvements}
To compute the limits  on the maximum  allowed  number of events we have used  the so called {\it optimum interval method} to find
an upper limit for a one-dimensionally distributed signal in the presence of an unknown background
\cite{Yellin:2002xd}.
The expected signal is computed by Montecarlo.
In the Montecarlo,  particles are extracted on a cylindrical  surface much larger than the
antenna bar. 
The Montecarlo therefore computes the acceptance in the case of simulated events that release at least 0.1 K and survive the analysis cuts. 

\begin{figure}[h]
\includegraphics[width=8.5cm]{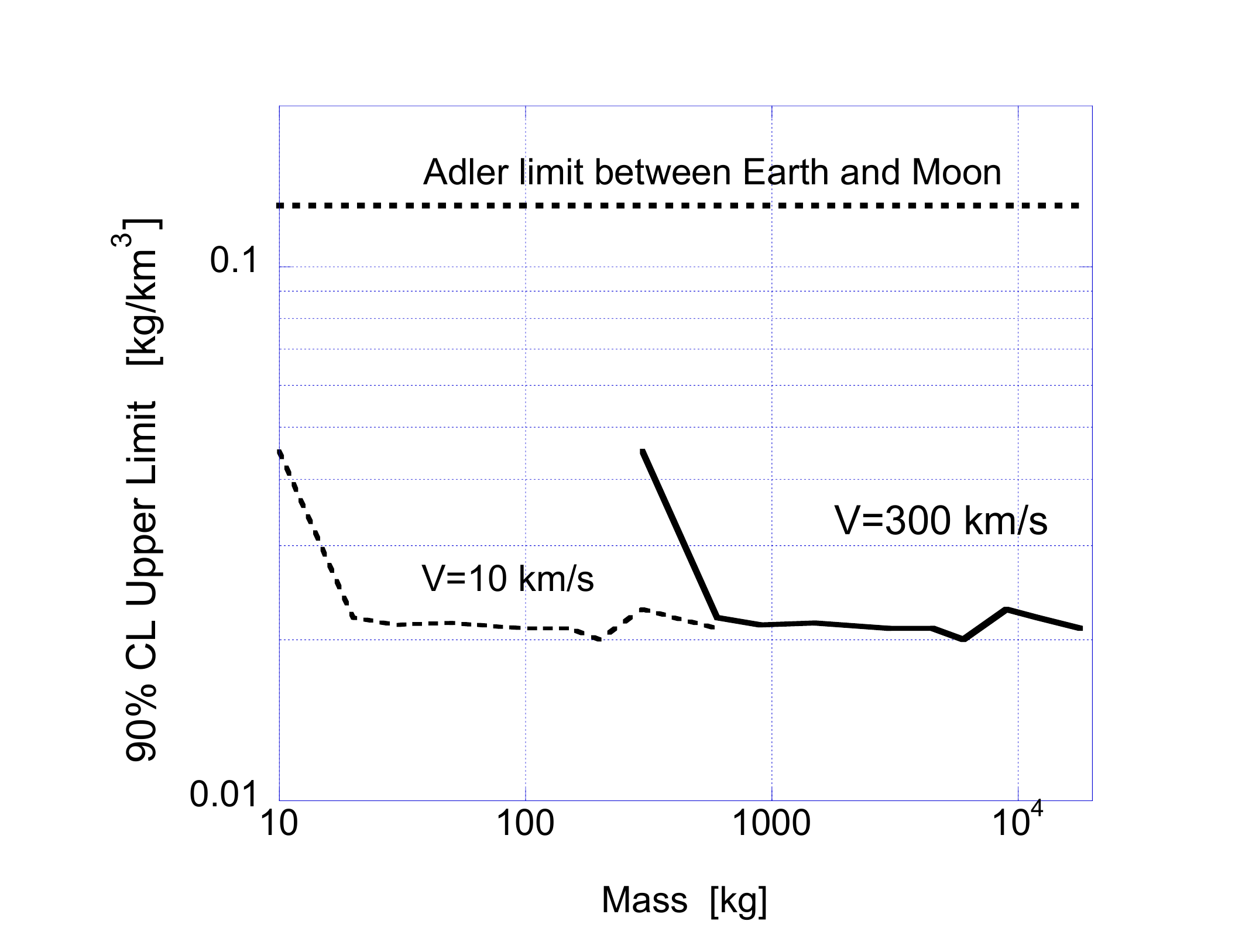}
\begin{minipage}[b]{7cm}\caption{\label{limitiNewt2}Newtorites density upper limits for v=10 and v=300 km/s vs  the Newtorite
mass. The limits are obtained from the Nautilus 2011 data set. As the  mass increases the limit reaches a plateau.
Adler \cite{Adler:2008rq} obtains a  direct upper limit of the mass of Earth-bound dark
matter lying between the radius of the moon orbit and the geodetic satellite orbit}
\end{minipage}
\end{figure}

The limits are shown in  Tab.~\ref{tb6} together with the acceptance. The Nautilus 2011 data set is the one with the lowest noise and gives Newtoriteslimits  a factor 2 better than the full Nautilus data set; the full data set, including Explorer, is used in the limits for the $MACRO$ particles. The limits are also shown in fig.\ref{limitiNewt2} for two values of the velocity $v$ =10
and  300 km/s. 
Our limit, although very far from the DM expected density ($5\cdot10^{-13}~kg/km^3$),  could be of some
interest due to the lack of other experimental limits derived from the direct detection of DM
particles that only interact gravitationally and on the Earth. 

There are several limits obtained studying the motion of celestial body in the solar system.
For example Adler \cite{Adler:2008rq} obtains a  direct upper limit of the mass of Earth-bound dark
matter lying between the radius of the moon orbit and the geodetic satellite orbit. The value
obtained is $0.13~ kg/km^3$, larger than our limit shown in fig.\ref{limitiNewt2}. 
Considering larger volumes  Pitjev \cite{Pitjev:2013sfa} has found a limit for possible DM inside
the Earth-Sun orbits of the order of $1.4\cdot10^{-7} kg/km^3$.

Our direct limit on Newtorites  could be improved by orders of magnitude using two or more nearby
bar antennas in coincidence. The performances for Newtorites of two antenna in coincidence have been
studied by a Montecarlo simulation that uses as input the Nautilus 2011 data set (therefore assuming the same
performances of Nautilus 2011).  
In the Montecarlo we  assumed  two antennas, positioned 1.5 m apart, with uncorrelated noise.  Larger distances, up to tens of meters, can still produce a detectable signal, depending on the value of $M/v$.
The result of this study is that a gain of about 300 seems to be possible
in 10 years of operations with noise similar to that of Nautilus 2011. This gain  is not enough
to reach the Pitjev bound. To reach this bound  it
is necessary to  increase the number of antennas and to reduce their noise. We recall that the current
antenna noise is limited by technology and is far from the intrinsic quantum limit of this kind of
device $\Delta E= \hbar\omega_{0}= 6~10^{-31}$ joules. So  a large R\&D effort would be necessary to
approach this limit.

The possibility to detect Newtorites in the $gw$ interferometric detector has been discussed by V. Frolov in a talk at the
GWDAW 2014 conference in Takayama (Japan). The detection is based on the measurement of an acceleration  on the mirrors.
The best acceleration sensitivity in the aLigo interferometer optimized at low frequency  is around $10^{-15} m~s^{-2}~Hz^{-1/2}$ at 20Hz to be compared to  $10^{-18} m~s^{-2}~Hz^{-1/2}$  for a Newtorite mass of $\sim$1kg at a distance of 7 km. Study are in progress to evaluate the signal to noise ratio using signal templates. Much better prospects are for the planned ET interferometer having a better sensitivity and the possibility of coincidences between the three interferometers.

\begin{figure}[h]
\includegraphics[width=7cm]{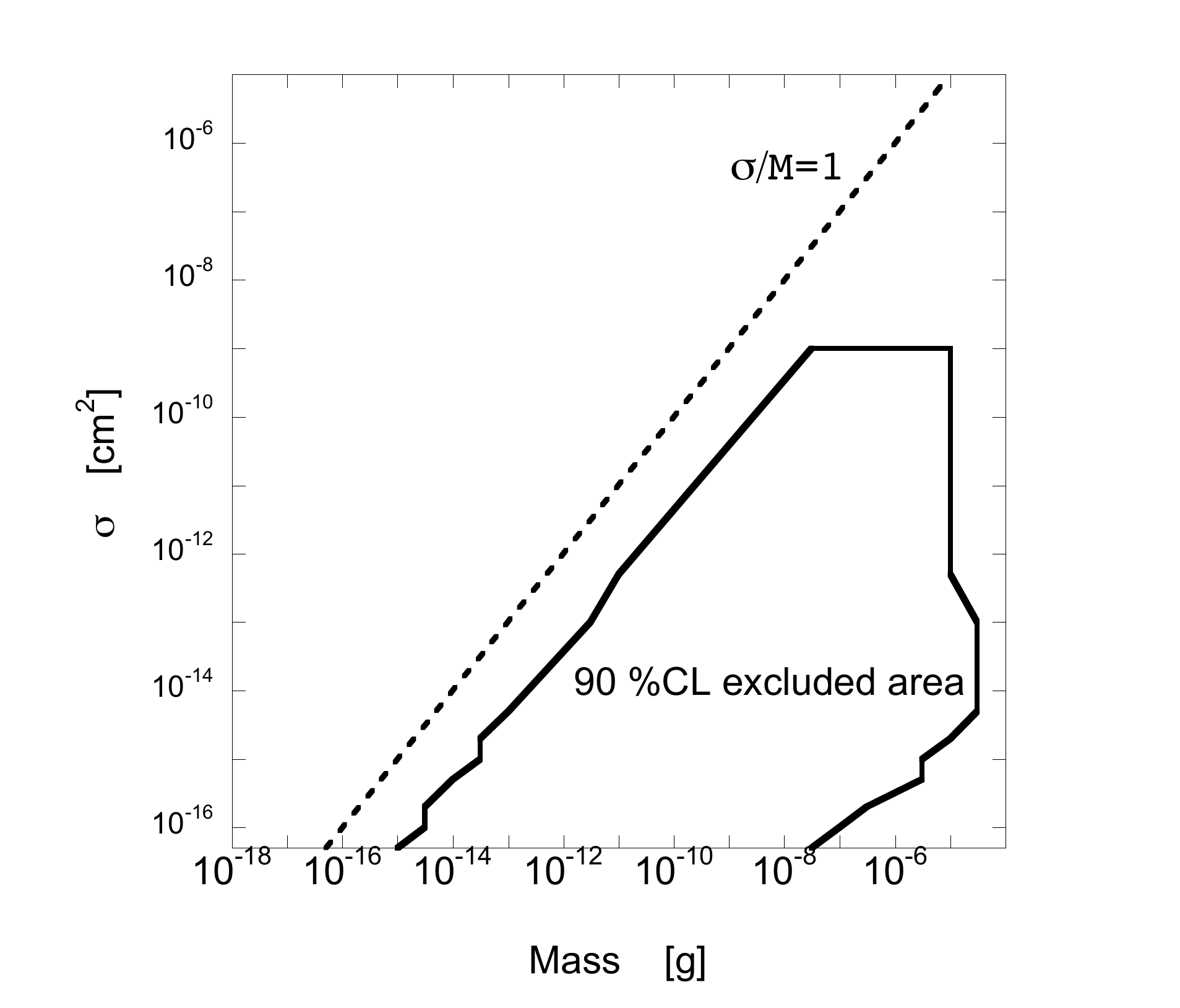}
\begin{minipage}[b]{8.7cm} \caption{\label{esclMACRO}{\it MACRO} particles 90 \% CL excluded regions in the plane cross section - mass, computed for $\beta=10^{-3}$ particles. The  line $\sigma$/M=1 is drawn for reference.}
\end{minipage}
\end{figure}

Finally  it's important to note that interesting limits can be obtained with the $gw$ bar data for other exotic particles, like the {\it MACRO} particles. An allowed region for the {\it MACRO} dark matter can be obtained in the plane cross section  vs. mass, by
excluding the region with upper limit on the flux smaller than the dark matter flux, following the
approach of Ref~\cite{Jacobs:2015csa}.
Fig.\ref{esclMACRO} shows the excluded region. It is interesting to observe that studies of
cosmological galaxy and cluster halos suggest a value $\sigma/M=0.1$ $cm^2$/g. 
For the comparison with constraints using other techniques see ref.~\cite{Jacobs:2015csa}.

\section*{References}

\begin{thebibliography}{}


\bibitem{Astone:2013bed} 
 P.~Astone {\it et al.},
 Proceedings of the 33Rd International Cosmic Ray Conference, Rio De Janeiro 2013
 arXiv:1306.5164 [astro-ph.HE].
 
 \bibitem{deru} C.~Bernard, A.~De Rujula, and B.~Lautrup, Nucl.\ Phys.\ B 242 (1984) 93.

\bibitem{Jacobs:2015csa} 
  D.~M.~Jacobs, G.~D.~Starkman and A.~Weltman,
  Phys. \ Rev.\  D {\bf91}, 115023 (2015)

\bibitem{nautilus} P.~Astone {et al.}  {\it Astropart. Phys.} { 7} (1997) 231.

\bibitem{Yellin:2002xd} 
  S.~Yellin,
  Phys.\ Rev.\ D {\bf 66}, 032005 (2002)
  [physics/0203002].
  %

\bibitem{Adler:2008rq} 
  S.~L.~Adler,
  J.\ Phys.\ A {\bf 41}, 412002 (2008)
  [arXiv:0808.0899 [astro-ph]].
  
\bibitem{Pitjev:2013sfa} 
  N.~P.~Pitjev and E.~V.~Pitjeva,
  Astron.\ Lett.\  {\bf 39}, 141 (2013)
  [Astron.\ Zh.\  {\bf 39}, 163 (2013)]
  %

\end{thebibliography}
\end{document}